\documentclass[fp,twocolumn]{jpsj3}
\def\Vec#1{\mbox{\boldmath $#1$}}

\title{Decoupling between Field-instabilities of Antiferromagnetism and Pseudo-metamagnetism in Rh-doped CeRu$_2$Si$_2$ Kondo Lattice}

\author{
Dai~{\sc Aoki}$^1$\thanks{E-mail address: dai.aoki@cea.fr}, 
Carley~{\sc Paulsen}$^2$,
Hisashi~{\sc Kotegawa}$^{1,3}$,
Fr\'{e}d\'{e}ric~{\sc Hardy}$^4$,
Christoph~{\sc Meingast}$^4$,
Pierre~{\sc Haen}$^2$,
Mounir~{\sc Boukahil}$^1$, 
William~{\sc Knafo}$^5$,
Eric~{\sc Ressouche}$^1$,
Stephane~{\sc Raymond}$^1$, and
Jacques~{\sc Flouquet}$^1$\thanks{E-mail address: jacques.flouquet@cea.fr}
}

\inst{%
$^1$INAC, SPSMS, CEA Grenoble, 38054 Grenoble, France\\
$^2$Institut N\'{e}el, CNRS/UJF Grenoble, 38042 Grenoble, France. \\
$^3$Department of Physics, Kobe University, Kobe 657-8501, Japan \\
$^4$Karlsruher Institut f\"{u}r Technologie, Institut f\"{u}r Festk\"{o}rperphysik, 76021 Karlsruhe, Germany\\
$^5$Laboratoire National des Champs Magn\'{e}tiques Intenses, UPR 3228, CNRS-UJF-UPS-INSA, 143 Avenue de Rangueil, 31400 Toulouse, France
}

\abst{
Doping Kondo lattice system CeRu$_2$Si$_2$ with Rh-$8\,{\%}$ (Ce(Ru$_{0.92}$Rh$_{0.08}$)$_2$Si$_2$) leads to drastic consequences 
due to the mismatch of the lattice parameters between CeRu$_2$Si$_2$ and CeRh$_2$Si$_2$.
A large variety of experiments clarifies the unusual properties of the ground state 
induced by the magnetic field from longitudinal antiferromagnetic (AF) mode at $H=0$ 
to polarized paramagnetic phase in very high magnetic field.
The separation between AF phase, paramagnetic phase and polarized paramagnetic phase varying with temperature, magnetic field and pressure is discussed on the basis of the experiments down to very low temperature.
Similarities and differences between Rh and La substituted alloys are discussed with 
emphasis on the competition between transverse and longitudinal AF modes,
and ferromagnetic fluctuations.
}

\kword{CeRu$_2$Si$_2$, Rh-doping, La-doping, antiferromagnetism, pseudo-metamagnetism, specific heat, resistivity, thermal expansion, magnetization, neutron diffraction, high pressure}

\begin{document}
\maketitle

\section{Introduction}
The effect of pressure on Ce heavy fermion systems close to the antiferromagnetic (AF) quantum phase
transition is well known.
Pressure drives the system from AF to paramagnetic (PM) ground state at a critical pressure $P_{\rm c}$.
If the transition is of second order with continuous suppression of the sublattice magnetization at $P_{\rm c}$,
$P_{\rm c}$ corresponds to a quantum critical point.~\cite{Flo05}

In complex materials such as heavy fermion compounds,
quite different pictures can be obtained for different materials as demonstrated in CeRu$_2$Si$_2$, CeCu$_6$ and YbRh$_2$Si$_2$.
The first case is often referred as an example of global criticality
where critical fluctuations are that of the magnetic order parameter, 
and two others of local criticality where quantum criticality is driven by local i. e.
$Q$-independent magnetic fluctuations~\cite{Flo05,Loh07}.
We focus here on the CeRu$_2$Si$_2$ series.
In this tetragonal crystal, the Ce ions show a strong Ising character with an anisotropy of susceptibility 
between the easy-axis ($c$-axis) and the hard-axis with a factor of $15$ at low temperatures.
The effect of the volume change via a tuning parameter $\delta$
have been achieved first by pressure or by doping on Ce$_{1-x}$La$_x$Ru$_2$Si$_2$.~\cite{Hol95,Flo02_CeRu2Si2}
The pure system CeRu$_2$Si$_2$ at $P=0$ is already in a PM ground state;
the effective critical pressure $P_{\rm c}$ is at a slightly negative pressure of a few kbar.
Expanding the volume by La substitution on the Ce side pushes to reenter in the AF domain 
for $x > x_{\rm c} \sim 0.075$;
the AF--PM transition at $T=0\,{\rm K}$ corresponds to a critical volume $V_{\rm c}$ achieved at $\delta_{\rm c}$
which is equivalent to $P_{\rm c}$ or $x_{\rm c}$.~\cite{Hae96,Que88}

Applying a magnetic field ($H$) will lead to cross the AF boundary $H_{\rm c}(T)$ between AF and PM.
For the Ising spin system of the CeRu$_2$Si$_2$ series
the first order metamagnetic transition at $H_{\rm c}(0)$ below $P_{\rm c}$ terminates at the critical endpoint $H_{\rm c}^\ast$.~\cite{Flo05,Flo02_CeRu2Si2,Flo10,Wei10}
The magnetic field can lead to switch from dominant AF interactions at low field
to a highly polarized paramagnetic phase (PPM) at high field
via a strong interplay between AF and ferromagnetic (FM) coupling through $H_{\rm c}$
studied with La- and Ge-doped system.~\cite{Flo05,Hae96,Hae99}
Furthermore, 
due to the large uniform susceptibility $\chi_0$ associated to the huge value of the Sommerfeld coefficient $\gamma$
(directly linked to the strong local 4\textit{f} character of the heavy fermion quasiparticle via Kondo fluctuation),
the growth of the majority spin-up component at $H_{\rm c}$ is associated to a Fermi surface instability.~\cite{AokiH95,Jul94,Dao06,Miy06}
Roughly at $T=0\,{\rm K}$, the magnetic polarization is given by the ratio of the induced magnetization ($\chi_0 H$)
by the saturated magnetization.
It reaches a critical value, $M_{\rm c}\sim 0.6\,\mu_{\rm B}/{\rm Ce}$ for $H_{\rm c} \sim H_{\rm K}$ Kondo field associated to the local spin fluctuations.

For $P>P_{\rm c}$, 
the first order metamagnetic transition is replaced 
by crossover phenomena at $H_{\rm m}$ referred as a pseudo-metamagnetism.
The effect can be quite sharp since close to $P_{\rm c}$, Kondo fluctuation and AF intersite interaction 
have comparable strength.
Approaching $P_{\rm c}$,
the crossover field $H_{\rm m}\sim H_{\rm K}$ is reduced and 
is comparable to $H_{\rm c}$ at $P_{\rm c}$ for $T=0\,{\rm K}$.

\begin{figure}[tbh]
\begin{center}
\includegraphics[width=1 \hsize,clip]{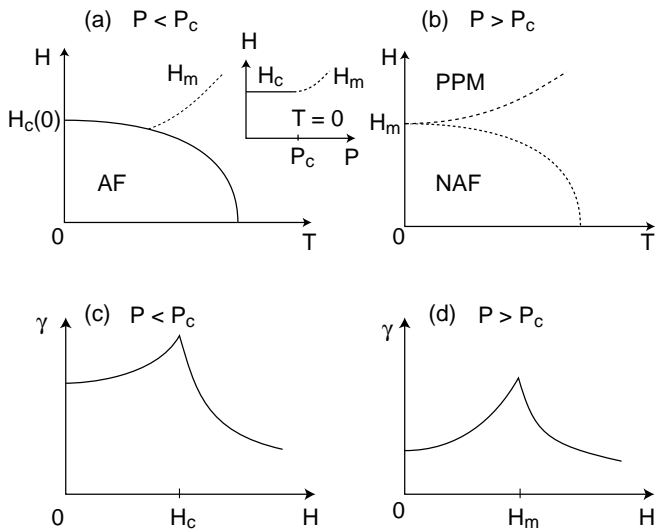}
\end{center}
\caption{Interplay between AF and pseudo-metamagnetism for (a) $P<P_{\rm c}$ and (b) $P>P_{\rm c}$.
The inset of panel (a) shows the pressure variation of $H_{\rm c}$ and $H_{\rm m}$ at $T=0\,{\rm K}$.
Field variation of the Sommerfeld coefficient $\gamma(H)$ for (c) $P<P_{\rm c}$ and (d) $P>P_{\rm c}$.
AF, PPM and NAF denote antiferromagnetism, polarized paramagnetism and nearly antiferromagnetism, respectively.}
\label{fig:phase_gamma_schematic}
\end{figure}

The combination of pressure and magnetic field gives the opportunity to observe 
the interplay between $H_{\rm c}(T,P)$ and $H_{\rm m}(T,P)$ schematically shown in Figs.~\ref{fig:phase_gamma_schematic}(a) and \ref{fig:phase_gamma_schematic}(b)
assuming an unique AF instability at a wave vector $\Vec{k}_1$.
For $P < P_{\rm c}$ in the AF domain,
the $H_{\rm m}$ crossover line joins the $H_{\rm c}(T)$ one at finite temperature.
For $P > P_{\rm c}$ the $H_{\rm m}$ line marks the entrance in the PPM region.
Measuring thermal expansion or specific heat for $P > P_{\rm c}$ shows that
two main clear crossover regimes delimited by a $\tilde{T}(H)$ line emerge:~\cite{Lac89,Pau90}
a low field nearly AF (NAF) regime dominated by AF correlations and
a high field one (PPM) above $H_{\rm m}$ governed by the strong local polarization of the electron.
Just in the vicinity of $H_{\rm c}$ or $H_{\rm m}$ strong duality between AF at $\Vec{k}_1$ and FM fluctuations at $k=0$ fluctuations is observed~\cite{Ros88,Flo04,Sat04}.
The expected variation of $\gamma(H)$ for both cases are shown in Figs.~\ref{fig:phase_gamma_schematic}(c) and \ref{fig:phase_gamma_schematic}(d).
Sharp maxima of $\gamma$ are expected at $H_{\rm c}$ and $H_{\rm m}$ on both sides
of $P_{\rm c}$ in agreement with the theoretical framework on AF tricriticality.
However, the first order nature of the metamagnetic transition can wipe out the sharp increase of $\gamma$
on approaching $H_{\rm c}$.

Furthermore in this complex band structure case, the magnetic field can induce changes in the ordered AF wave vector 
as often different magnetic hot spots exist on the Fermi surface.
Thus the selection of the AF wave vector can be modified by the magnetic field 
in the AF domain.

Inelastic neutron scattering experiments on the pure CeRu$_2$Si$_2$ system indicate 
that three AF hot spots exist at $\Vec{k}_1=(0.31,0,0)$, $\Vec{k}_2=(0.31,0.31,0)$ and $\Vec{k}_3=(0,0,0.35)$.~\cite{Kad04}
The two first $\Vec{k}$ vectors, $\Vec{k}_1$ and $\Vec{k}_2$ are transverse modes as the fluctuating moments
are aligned along the $c$-axis of the tetragonal crystal,
while $\Vec{k}_3$ is a longitudinal one.
Under magnetic field, a sharp pseudo-metamagnetic crossover occurs at $H_{\rm m}$ marked by a sharp effective mass enhancement right at $H_{\rm m}$, 
a slowdown of the field induced FM fluctuation, and 
a spectacular Fermi surface change.~\cite{Flo05,Flo02_CeRu2Si2,Flo04,Sat04,AokiH95,Jul94}

\begin{figure}[tbh]
\begin{center}
\includegraphics[width=1 \hsize,clip]{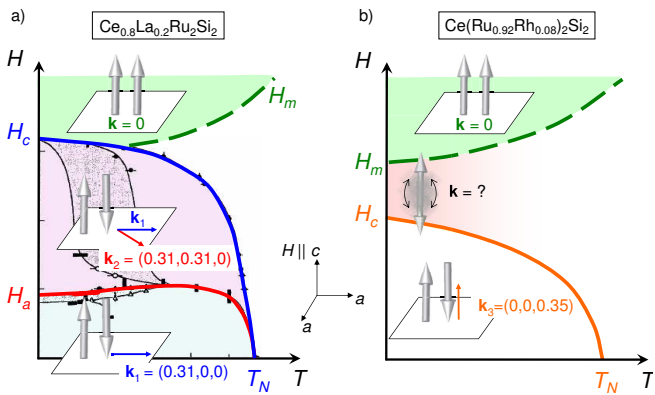}
\end{center}
\caption{(Color online) Schematic magnetic phase diagrams of Ce$_{0.8}$La$_{0.2}$Ru$_2$Si$_2$ and Ce(Ru$_{0.92}$Rh$_{0.08}$)$_2$Si$_2$.}
\label{fig:phase_schematic}
\end{figure}

For the La substituted case, AF ordering at $H=0$ occurs at a transverse wave vector $\Vec{k}_1$
with respect to the sublattice magnetization aligned along $c$-axis.~\cite{Que88}
At low temperatures, the sharp pseudo-metamagnetic crossover at $H_{\rm m}$ is replaced by
a first order metamagnetic transition at $H_{\rm c}$.
Increasing temperature, the field sweep below the N\'{e}el temperature $T_{\rm N}$ leads to 
detect both $H_{\rm c}(T)$ and $H_{\rm m}(T)$.
At $x_{\rm c}$ for $T \to 0\,{\rm K}$, $H_{\rm c}$ is equal to $H_{\rm m}$.
The ($H,T$) phase diagram is more complex
than that shown in Fig.~\ref{fig:phase_gamma_schematic} as at a critical field $H_{\rm a} < H_{\rm c}$,
the AF ordered wave vector remains transverse but changes from incommensurate wave vector to another complex AF phase 
with a large hysteresis domain characterized on cooling by the appearance of the wave vector $\Vec{k}_2$
and even a commensurate component $(1/3,1/3,0)$ at low temperature.~\cite{Mig91,Mig90}
For $x$ larger than $x_{\rm c}$, the schematic scheme of the ($H,T$) phase diagram measured by
the elastic neutron diffraction is shown in Fig.~\ref{fig:phase_schematic}(a).

In the previous studies with La substitution, the $H_{\rm c}(T)$ and $H_{\rm m}(T)$ phase diagram interfere,
leading to an AF order to PPM phase boundary at $H_{\rm c}$ for $\delta < \delta_{\rm c}$.
There is no obvious basic arguments for the coincidence between $H_{\rm c}$ and $H_{\rm m}$ at $\delta_{\rm c}$.
However, it is clear that the dominant competing wave vector are the AF transverse ones ($\Vec{k}_1$, $\Vec{k}_2$, \ldots) 
and the uniform FM mode.

A striking point is that Rh substitution on the Ru site, namely Ce(Ru$_{1-x}$Rh$_x$)$_2$Si$_2$ reveals a decoupling between $H_{\rm c}$ and $H_{\rm m}$.~\cite{Sek92,Sek93,Sak92}
This drastic change may be driven by the fact that
with Rh doping the ordered wave vector in the AF domain ($x>x_{\rm c} = 0.05$)
is now the longitudinal ($\Vec{k}_3=(0,0,0.35)$) wave vector instead of the transverse mode in the La doped alloys (Fig.~\ref{fig:phase_schematic}(b)).~\cite{Kaw97,Kad06}
Up to now there is no indication via neutron scattering experiments what will be the dominant AF spin fluctuation mode up to $H_{\rm m}$.

Previous experiments on Ce(Ru$_{1-x}$Rh$_x$)$_2$Si$_2$ were limited at temperatures above $1.5\,{\rm K}$ except for $x=0.15$ which is quite higher than $x_{\rm c}\sim 0.05$.~\cite{Sek97_CeRu2Si2,Sek98}
Thus we performed new set of experiments on $x=0.08$.
To clarify the situation, we focus here on the thermodynamic and transport measurements at very low temperatures down to $100\,{\rm mK}$.
The first aim is to determine the field variation of the Sommerfeld coefficient $\gamma$ through $H_{\rm c}$ and $H_{\rm m}$
and to compare the results with magnetization, specific heat and resistivity measurements.
The second aim is to investigate precisely the crossover line $\tilde{T}(H)$ by high accurate thermal expansion measurements in the intermediate temperature range above $2\,{\rm K}$.
The third aim is to clarify the pressure evolution of $H_{\rm m}$ and the effective mass, comparing with those of CeRu$_2$Si$_2$.

It will be shown that the substitution of Ru by Rh is a major perturbation due to the mismatch of the lattice parameter of CeRu$_2$Si$_2$
and CeRh$_2$Si$_2$.
In the extended field region $H_{\rm m}-H_{\rm c} = 3\,{\rm T}$ assumed to be paramagnetic,
the fancy point is the quasi-invariant value of $\gamma$ close to the critical value $\gamma_{\rm c}$ observed at $H_{\rm c}^\ast$
or $P_{\rm c}$ at $H=0$ in the La doped system.
The validity of the hypothesis of PM ground state above $H_{\rm c}$ for longitudinal and transverse modes will be confirmed by new neutron scattering experiments.

\section{Experimental}
Single crystals of Ce(Ru$_{0.92}$Rh$_{0.08}$)$_2$Si$_2$ were grown using the Czochralski method in a tetra-arc furnace.
Starting materials of Ce (purity: 99.9{\%}-3N), Ru(4N), Rh(4N) and Si(6N) with the ratio, 1 : 1.84 : 0.16 : 2
were melted under the high purity Ar gas for a polycrystalline ingot.
The ingot was tuned over and was melted again. 
This process was repeated several times in order to obtain the homogeneous phase.
The ingot was subsequently pulled with a pulling rate of $15\,{\rm mm/hr}$.
The obtained single crystal ingot was cut using a spark cutter, and was oriented by the X-ray Laue photograph.
The first specific heat measurements were realized using the relaxation method 
at temperatures down to $0.42\,{\rm K}$ and at magnetic fields up to $9\,{\rm T}$.
The measurements were then pushed down to $0.21\,{\rm K}$, using a homemade dilution refrigerator up to $14\,{\rm T}$.
Precise magnetization measurements were performed down to $80\,{\rm mK}$ and up to $8\,{\rm T}$ with a homemade SQUID magnetometer,
which was successfully applied in our previous experiments CeRu$_2$Si$_2$~\cite{Pau90} and CeCoIn$_5$~\cite{Pau11}.
Thanks to the Maxwell relation, $\partial \gamma /\partial H = \partial^2 M/\partial T^2$,
the field dependence of $\gamma$ can be determined by integrating $\partial^2 M/\partial T^2$ with field.
Resistivity measurements were performed using four probe AC method at temperature down to $100\,{\rm mK}$ and at field up to $16\,{\rm T}$.
The pressure study was also realized by the resistivity measurements in a NiCrAl-CuBe pressure cell up to $3.8\,{\rm kbar}$. 
The pressure was determined by the superconducting transition temperature of Pb.
The $H_{\rm c}(T)$ and $H_{\rm m}(T)$ boundaries were confirmed by magnetostriction experiments 
using a strain gauge glued on the $c$-plane down to $2\,{\rm K}$ and up to $9\,{\rm T}$.
The search for $\tilde{T}(H)$ line was precisely investigated by high precision thermal expansion measurements using a capacitance dilatometer 
down to $2\,{\rm K}$.
For comparison, thermal expansion measurements were performed in Ce$_{0.9}$La$_{0.1}$Ru$_2$Si$_2$, as well.
The single crystals obtained in the previous reports~\cite{Fis91,Aok11_CeRu2Si2} were used.
The precision of our measurements has been improved, 
compared with the previous works realized down to $1.2\,{\rm K}$.~\cite{Bio99,Hae01}
In order to confirm that $H_{\rm c}$ marks the AF--PM boundary,
a neutron diffraction experiment was performed on the two-axis D23-CRG-CEA thermal neutron diffractometer 
equipped with a lifting detector at ILL in Grenoble.
A copper monochromator provides an unpolarized neutron beam with a wavelength of $\lambda=1.276\,{\rm \AA}$.
The single crystal sample was put in the $12\,{\rm T}$ vertical field magnet 
with the $c$-axis along the field.

\section{Results}
\subsection{Ce(Ru$_{0.92}$Rh$_{0.08}$)$_2$Si$_2$}
\begin{figure}[tbh]
\begin{center}
\includegraphics[width=1 \hsize,clip]{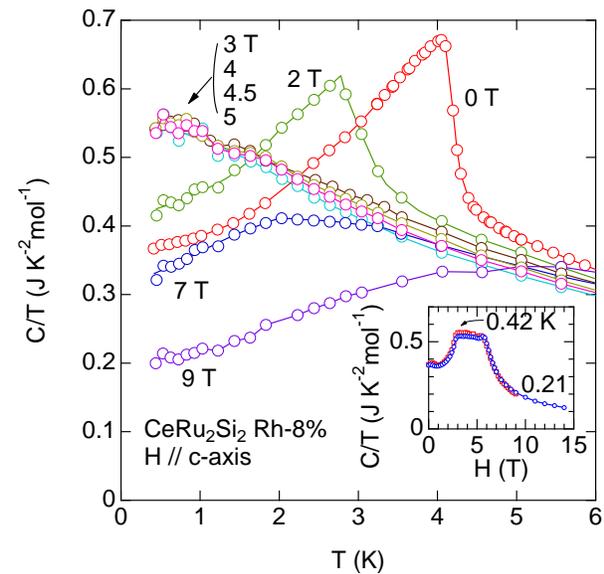}
\end{center}
\caption{(Color online) Temperature dependence of the specific heat in the form of $C/T$ vs $T$ at different fields 
in Ce(Ru$_{0.92}$Rh$_{0.08}$)$_2$Si$_2$. The inset shows the field dependence of the specific heat in the form of $C/T$ vs $H$ at $0.21\,{\rm K}$ and $0.42\,{\rm K}$.}
\label{fig:Cp_Tdep}
\end{figure}

Figure~\ref{fig:Cp_Tdep} shows the temperature dependence of $C/T$ at different magnetic fields.
AF transition occurs at $T_{\rm N}=4.2\,{\rm K}$ at zero field.
As for classical AF, $T_{\rm N}$ identified as a jump of $C/T$ decreases with $H$.
That allows to determine $H_{\rm c}(0)\sim 2.8\,{\rm T}$.
Surprisingly between $3\,{\rm T}$ and $5.8\,{\rm T}$, $C/T$ is almost invariant against field
in a regime assumed to be paramagnetic.
At $H_{\rm m}\sim 5.8\,{\rm T}$, $C/T$ abruptly drops as observed for CeRu$_2$Si$_2$
just above the pseudo-metamagnetic field $H_{\rm m}\sim 7.8\,{\rm T}$.

\begin{figure}[tbh]
\begin{center}
\includegraphics[width=1 \hsize,clip]{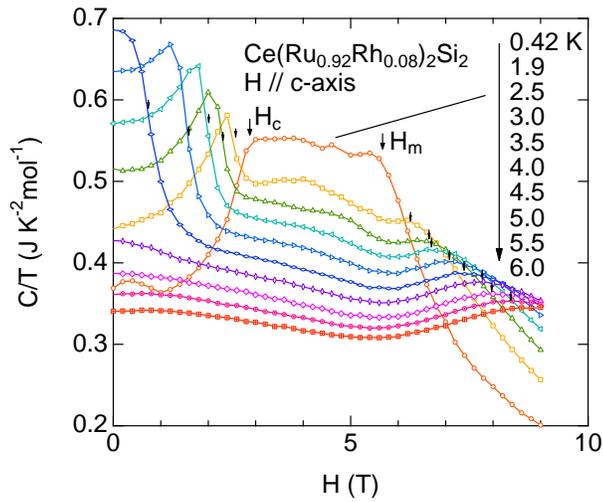}
\end{center}
\caption{(Color online) Field dependence of the specific heat in the form of $C/T$ vs $H$ at different temperatures in Ce(Ru$_{0.92}$Rh$_{0.08}$)$_2$Si$_2$. The small arrows indicate $H_{\rm c} (T)$ and $H_{\rm m}(T)$.}
\label{fig:Cp_Hdep}
\end{figure}

Figure~\ref{fig:Cp_Hdep} shows the field dependence of $C/T$ at different temperatures
with again the field decrease of the AF anomaly at $T_{\rm N}(H)$ and
the smearing out of the PPM domain on warming at $H_{\rm m}(T)$.

\begin{figure}[tbh]
\begin{center}
\includegraphics[width=1 \hsize,clip]{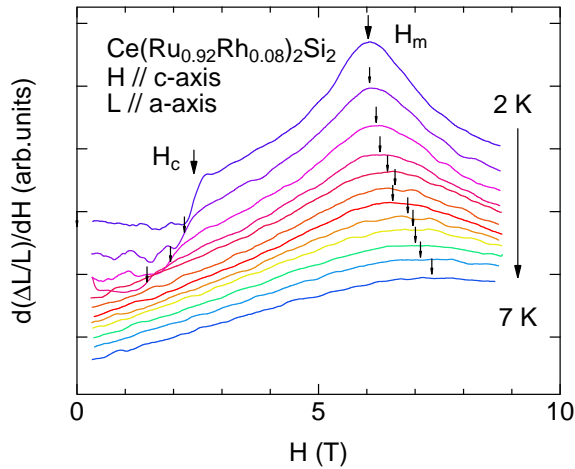}
\end{center}
\caption{(Color online) Field dependence of the magnetostriction in the form of $d (\Delta L/L)/dH$ vs $H$ at different fields in Ce(Ru$_{0.92}$Rh$_{0.08}$)$_2$Si$_2$.}
\label{fig:MS_Hdep}
\end{figure}

\begin{figure}[tbh]
\begin{center}
\includegraphics[width=1 \hsize,clip]{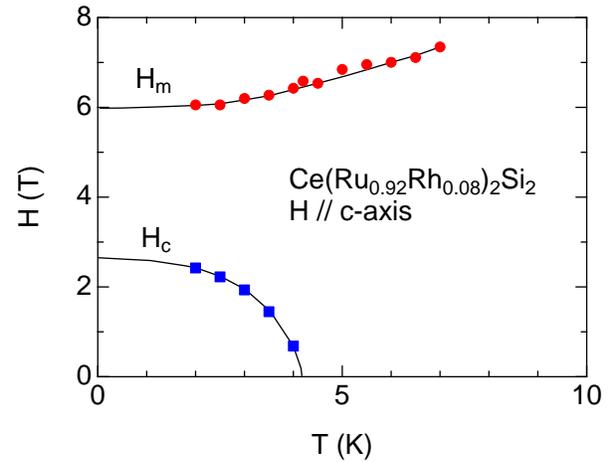}
\end{center}
\caption{(Color online) ($H,T$) phase diagram obtained by the magnetostriction measurements in Ce(Ru$_{0.92}$Rh$_{0.08}$)$_2$Si$_2$.}
\label{fig:MS_HT_phase}
\end{figure}

Figure~\ref{fig:MS_Hdep} represents the response of the strain gauge directly linked with the change of the length
$\Delta L/L$ along the $c$-axis.
Taking the field derivative $\lambda_c = d (\Delta L/L)/dH$,
the positions of the lines $H_{\rm c}(T)$ and $H_{\rm m}(T)$ are well drawn (Fig.~\ref{fig:MS_HT_phase}).
The singular point is that the magnetostriction at $H_{\rm m}$ appears quite more broadened than 
the one observed for CeRu$_2$Si$_2$ at the PM--PPM boundary.~\cite{Flo05,Hol95,Flo02_CeRu2Si2,Lac89,Pau90} 
As discussed later,
this smearing is caused by the doping which inhibits partly a full deformation of the lattice.

\begin{figure}[tbh]
\begin{center}
\includegraphics[width=1 \hsize,clip]{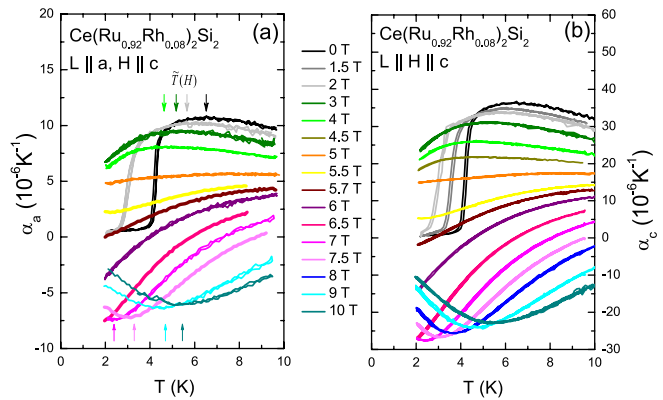}
\end{center}
\caption{(Color online) Thermal expansion of Ce(Ru$_{0.92}$Rh$_{0.08}$)$_2$Si$_2$ with high accuracy at the intermediate temperature regime ($T>2\,{\rm K}$) at different fields for $L\parallel a$-axis and $L\parallel c$-axis.}
\label{fig:Texp_Rh}
\end{figure}

\begin{figure}[tbh]
\begin{center}
\includegraphics[width=0.8 \hsize,clip]{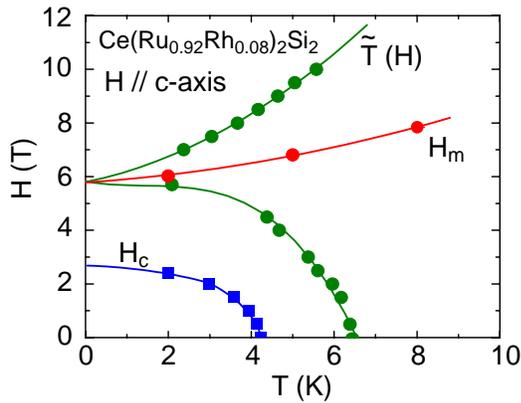}
\end{center}
\caption{(Color online) ($H,T$) phase diagram of Ce(Ru$_{0.92}$Rh$_{0.08}$)$_2$Si$_2$ determined by the precise thermal expansion measurements.}
\label{fig:Texp_phase_Rh}
\end{figure}

\begin{figure}[tbh]
\begin{center}
\includegraphics[width=1 \hsize,clip]{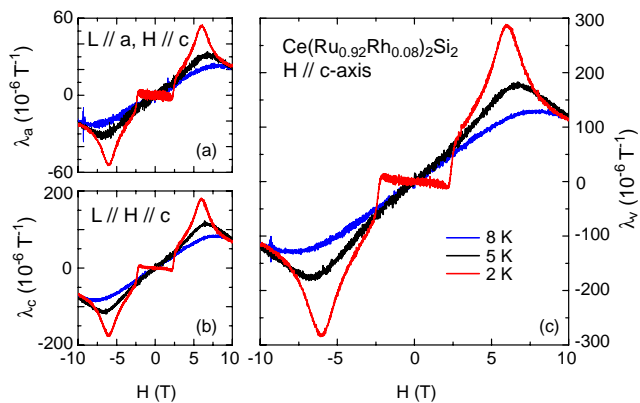}
\end{center}
\caption{(Color online) Field dependence of the magnetostriction $\lambda_V = (1/V)(\partial V/\partial H)$ on Ce(Ru$_{0.92}$Rh$_{0.08}$)$_2$Si$_2$.}
\label{fig:10}
\end{figure}

A new set of thermal expansion measurements $\alpha$ was realized to search for the crossover $\tilde{T}(H)$ line for $H\parallel c$, 
as shown in Fig.~\ref{fig:Texp_Rh}.
As observed in the La-doped case,~\cite{Lac88} the same sign of thermal expansion is detected along $c$ and $a$-axes with a ratio near 3.
The AF $H_{\rm c}(T)$ line is very well defined through the strong negative jump of $\alpha$.
The crossover line $\tilde{T}(H)$ can be drawn following the position of the extremum of thermal expansion at fixed field (Fig.~\ref{fig:Texp_phase_Rh}).
This indicates strongly that the region between $H_{\rm m}$ and $H_{\rm c}$ is paramagnetic.
The ($H,T$) phase diagram can be reconstructed with a mixture of phase diagram shown in Fig.~\ref{fig:phase_gamma_schematic}.
A clear separation exists between $H_{\rm c}$ and $H_{\rm m}$.
The effect is obvious in the magnetostriction data $\lambda_{\rm V}=(1/V)\partial V/\partial H$
shown in Fig.~\ref{fig:10} even for $T\sim 2\,{\rm K}$ equal only to $0.5 T_{\rm N}$.


\begin{figure}[tbh]
\begin{center}
\includegraphics[width=1 \hsize,clip]{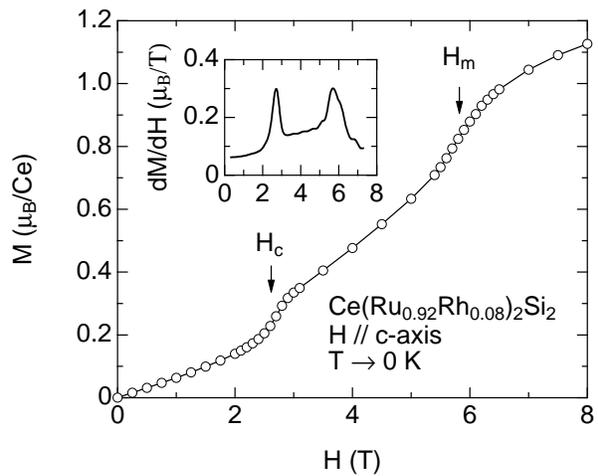}
\end{center}
\caption{Magnetization curve extrapolated to $0\,{\rm K}$ obtained by the temperature dependence at different fields in Ce(Ru$_{0.92}$Rh$_{0.08}$)$_2$Si$_2$. The inset shows the field derivative of the magnetization curve.}
\label{fig:mag}
\end{figure}

Figure~\ref{fig:mag} shows the magnetization curve extrapolated to $0\,{\rm K}$ 
from the temperature dependence down to $75\,{\rm mK}$ at different fields.
A weak hysteresis is detected only at $H_{\rm c}$, as it was reported on Ce(Ru$_{0.85}$Rh$_{0.15}$)$_2$Si$_2$.~\cite{Sak92}
Taking the $H$ derivative $\chi (H)=\partial M/\partial H$ of the magnetization (inset of Fig.~\ref{fig:mag}),
one can observe that the transition at $H_{\rm c}$ and $H_{\rm m}$ are very broad 
by comparison to the first order metamagnetic transition detected for example on Ce$_{0.9}$La$_{0.1}$Ru$_2$Si$_2$ ($T_{\rm N}=5\,{\rm K}$)
at $H_{\rm c}$ and 
to the pseudo-metamagnetic transition of CeRu$_2$Si$_2$ where $\chi (H_{\rm m})$ reaches $1.6\,\mu_{\rm B}/{\rm T}$
at $100\,{\rm mK}$.~\cite{Flo05,Fis91}
This reduction points out that the magnetostriction is strongly affected as the susceptibility at constant pressure ($\chi_{\rm P}^{}$)
is linked to the susceptibility at constant volume $\chi_{\rm V}^{}$ by the relation:
\begin{equation}
\chi_{\rm P}^{} = \chi_{\rm V}^{} +  \lambda_{\rm V}{}^2 \frac{V_0}{\kappa},
\end{equation}
where $V_0$ is the molar volume and $\kappa$ is the compressibility.
A quite similar attenuation is observed when CeRu$_2$Si$_2$ is doped with La and Y in $\chi_{\rm P}^{}(H_{\rm m})$
and $\lambda_{\rm V} = (1/V) \partial V/\partial H$.~\cite{Hol95}
It is interesting to notice that in Ce(Ru$_{0.92}$Rh$_{0.08}$)$_2$Si$_2$ the switch from PM to PPM phase occurs at a critical value $M_{\rm c}\sim 0.7\,\mu_{\rm B}$, just slightly higher than the one measured in CeRu$_2$Si$_2$.

\begin{figure}[tbh]
\begin{center}
\includegraphics[width=1 \hsize,clip]{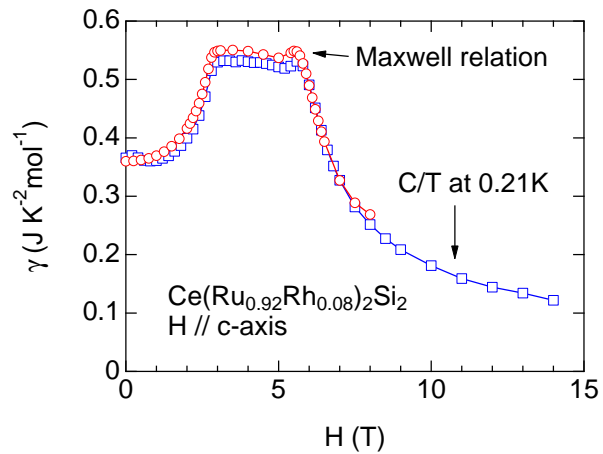}
\end{center}
\caption{(Color online) Field dependence of the Sommerfeld coefficient $\gamma$ obtained by the Maxwell relation from the magnetization measurements assuming $\gamma (0)=0.36\,{\rm J\, K^{-2} mol^{-1}}$. The result of direct specific heat measurements at $0.21\,{\rm K}$ is plotted as well for comparison.}
\label{fig:Maxwell}
\end{figure}

The absence of hysteresis between $H_{\rm c}$ and $H_{\rm m}$, the broad susceptibility anomaly at $H_{\rm m}$,
and the observation of crossover $\tilde{T}(H)$ line give strong supports that
the $H_{\rm m}$--$H_{\rm c}$ window corresponds to a paramagnetic ground state.
The paradox is that in this paramagnetic domain a quasi-constant field value of $\gamma$ is observed as derived with
great accuracy from the application of the Maxwell relation to the temperature dependence of the
magnetization at constant field, namely $\partial \gamma /\partial H = \partial^2 M/\partial T^2$ 
and also by direct specific heat measurements (Fig.~\ref{fig:Maxwell}).

\begin{figure}[tbh]
\begin{center}
\includegraphics[width=1 \hsize,clip]{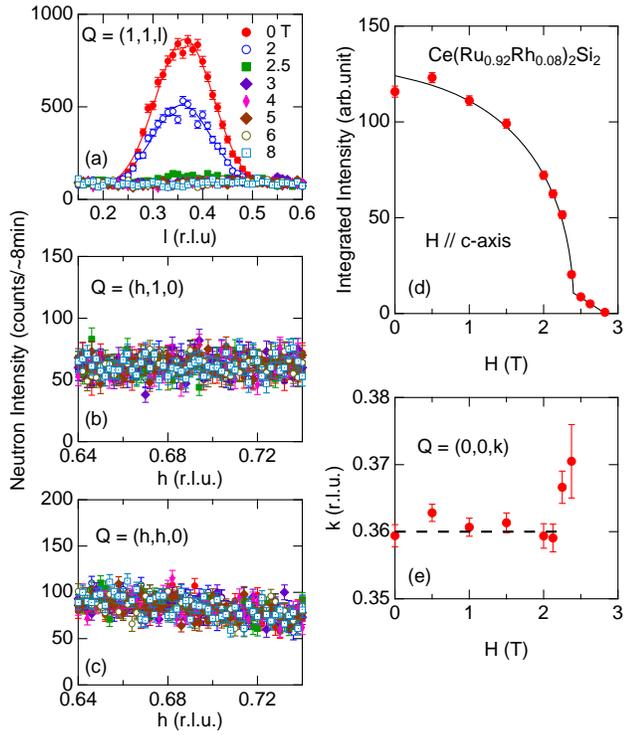}
\end{center}
\caption{(Color online) (a)--(c)Neutron diffraction profiles of  Ce(Ru$_{0.92}$Rh$_{0.08}$)$_2$Si$_2$ at $2.3\,{\rm K}$ at different fields for $H \parallel c$-axis along $(1,1,l)$, $(h,1,0)$ and $(h,h,0)$ corresponding to $\Vec{k}_3$, $\Vec{k}_1$ and $\Vec{k}_2$, respectively. (d)Field dependence of the integrated intensity of $(0,0,k)$ at $2.3\,{\rm K}$. (e)Corresponding field dependence of $k$.}
\label{fig:neutron}
\end{figure}

In order to establish that AF disappears at $H_{\rm c}$,
we have performed new elastic scattering experiments on a single crystal extracted from the same batch.
The aim was to verify the possible interplay between the longitudinal and transverse mode.
Figure~\ref{fig:neutron} shows the representative $\Vec{Q}$ scan performed around 
$\Vec{Q}=(1,1,0)+\Vec{k}_3$ (Fig.~\ref{fig:neutron}(a)),
$\Vec{Q}=(1,1,0)-\Vec{k}_1$ (Fig.~\ref{fig:neutron}(b)) and
$\Vec{Q}=(1,1,0)-\Vec{k}_2$ (Fig.~\ref{fig:neutron}(c)) at different field at $2.3\,{\rm K}$.
These data indicate that the magnetic ordering with the wave vector $\Vec{k}_3$ disappears at $H_{\rm c}$,
while no signal is induced at $\Vec{k}_1$ and $\Vec{k}_2$ for higher magnetic fields above $H_{\rm c}$ and $H_{\rm m}$.
Figure~\ref{fig:neutron}(d) shows the magnetic field dependence of the integrated intensity measured for $\Vec{k}_3$.
The value of $H_{\rm c}$ at $2.3\,{\rm K}$ is approximately $2.3\,{\rm T}$ in good agreement with the 
phase diagram drawn in Fig.~\ref{fig:Texp_phase_Rh}.
Short range correlations are observed up to $2.8\,{\rm T}$. 
Figure~\ref{fig:neutron}(e) gives the field dependence of the $l$ component of $\Vec{k}_3$ which is found to be $0.36$ at low field.

\begin{figure}[tbh]
\begin{center}
\includegraphics[width=0.8 \hsize,clip]{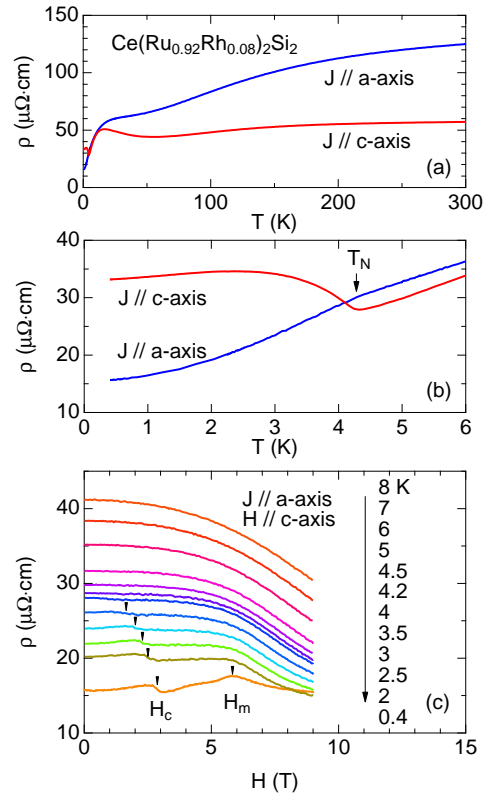}
\end{center}
\caption{(Color online) (a)Temperature dependences of the resistivities for the current along $a$ and $c$-axes and (b)the corresponding low temperature resistivities. (c)Magnetoresistance at different temperatures for $H\parallel c$-axis and $J\parallel a$-axis}
\label{fig:resist}
\end{figure}

\begin{figure}[tbh]
\begin{center}
\includegraphics[width=1 \hsize,clip]{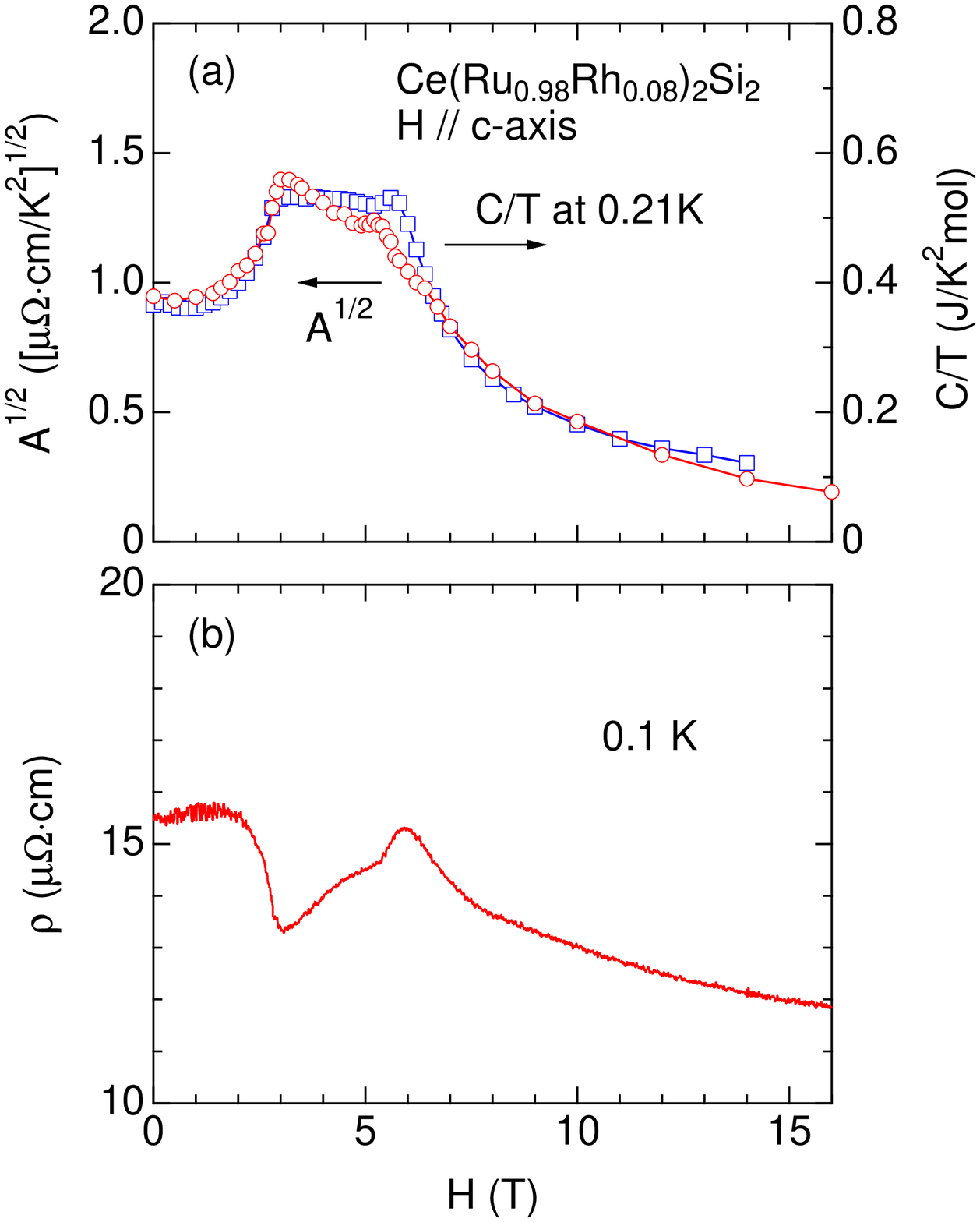}
\end{center}
\caption{(Color online) (a)Field dependence of the resistivity $A$ coefficient and the specific heat at $0.21\,{\rm K}$ in the form of $\sqrt{A}$ vs $H$ and $C/T$ vs $T$ in Ce(Ru$_{0.92}$Rh$_{0.08}$)$_2$Si$_2$. (b)Field dependence of the resistivity at $0.1\,{\rm K}$.}
\label{fig:A_rho0}
\end{figure}

Transport measurements were realized as shown in Figs.~\ref{fig:resist} and \ref{fig:A_rho0}.
The ($H,T$) phase diagram is well reproduced with a characteristic drop of the residual resistivity at $H_{\rm c}(0)$
followed by a wide maximum at $H_{\rm m}(T)$.
In agreement with the validity of the Kadowaki-Woods relation,
the $A$ coefficient of the $T^2$ resistivity law roughly follows the field dependence of $\gamma$
according to the relation $A\propto \gamma^2$.
In the case of doping, additional deviation from the Kadowaki-Woods relation can occur
due to the $T$ dependence of the impurity scattering and also due to the additional modification in the carrier number.~\cite{Kam97,Mat11_CeRu2Si2}

\begin{figure}[tbh]
\begin{center}
\includegraphics[width=1 \hsize,clip]{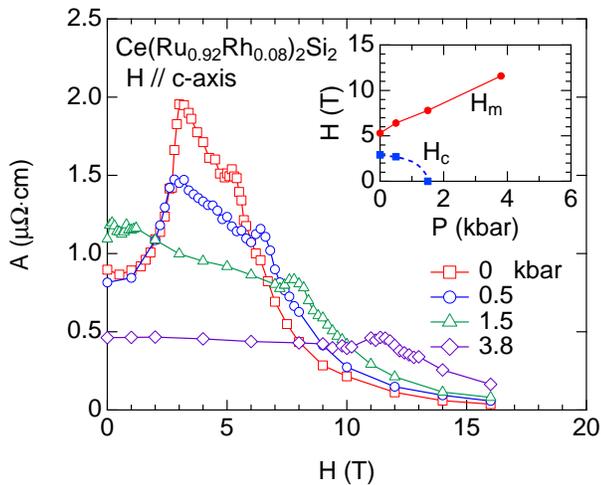}
\end{center}
\caption{(Color online) Field dependence of the $A$ coefficient of resistivity at different pressures for $H\parallel c$-axis in Ce(Ru$_{0.92}$Rh$_{0.08}$)$_2$Si$_2$.
The inset shows the pressure dependence of $H_{\rm m}$ and $H_{\rm c}$.}
\label{fig:resist_pressure}
\end{figure}

\begin{figure}[tbh]
\begin{center}
\includegraphics[width=1 \hsize,clip]{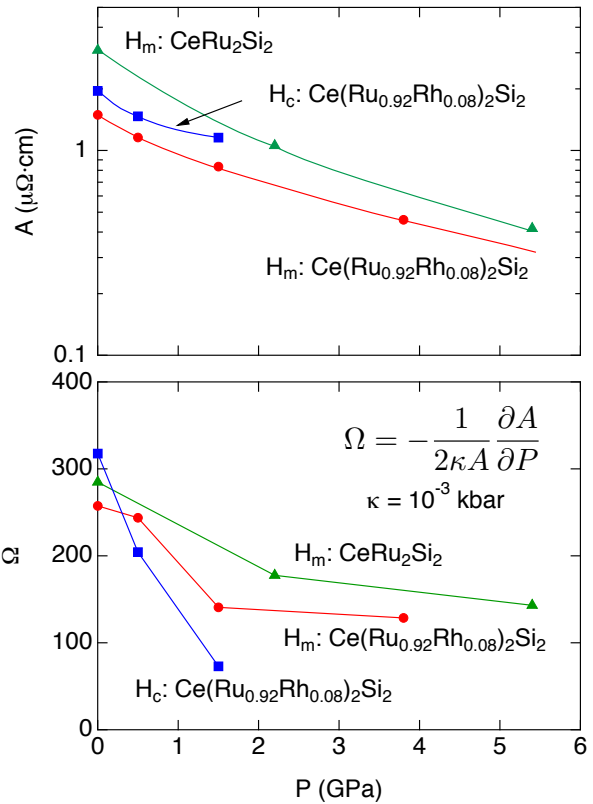}
\end{center}
\caption{(Color online) (a)Pressure dependence of the A coefficient at $H_{\rm m}$ and $H_{\rm c}$ in Ce(Ru$_{0.92}$Rh$_{0.08}$)$_2$Si$_2$ and CeRu$_2$Si$_2$.
(b)Pressure dependence of the effective Gr\"{u}neisen parameter $\Omega$ for $\sqrt{A}$ at $H_{\rm m}$ and $H_{\rm c}$ in CeRu$_2$Si$_2$ and Ce(Ru$_{0.92}$Rh$_{0.08}$)$_2$Si$_2$. $\Omega$ is defined as
$\Omega \equiv -\frac{\partial \ln m^\ast}{\partial \ln V}\sim -\frac{1}{2\kappa A}\frac{\partial A}{\partial P}$, where $\kappa$ and $m^\ast$ are compressibility and effective mass, respectively}
\label{fig:Gruneisen}
\end{figure}

Finally, preliminary pressure studies were realized to test the pressure response of $H_{\rm c}$ and $H_{\rm m}$
up to $3.8\,{\rm kbar}$, as shown in Fig.~\ref{fig:resist_pressure}.
With increasing pressure, 
the plateau of the resistivity $A$ coefficient between $H_{\rm m}$ and $H_{\rm c}$
decreases in value.
$H_{\rm c}$ disappears rapidly at $P\sim 1.5\,{\rm kbar}$, while $H_{\rm m}$ monotonously increases with pressure.
In this preliminary experiments, it is not possible to conclude 
that $H_{\rm c}$ terminates at $P_{\rm c}$ as a critical endpoint,
as it is in the La-doped case at $x_{\rm c}$.~\cite{Flo05}
A fine tuning through $P_{\rm c}$ is necessary.
Close to the critical pressure ($P\sim 1.5\,{\rm kbar}$)
where the longitudinal ordering collapses,
the critical value of $\gamma_{\rm c}$ may be quite lower than the value of 
$\gamma \sim 650\,{\rm mJ\,K^{-2}mol^{-1}}$ in Ce$_{0.9}$La$_{0.1}$Ru$_2$Si$_2$ reported for the AF instability 
of the transversal mode,~\cite{Fis91,Pue88,Aok11_CeRu2Si2}
as $A^{\rm Rh}(P_{\rm c})\sim 1.4 A^{\rm Rh}(P=0)$ at $H=0$ leading to $\gamma_{\rm c}\sim 450\,{\rm mJ\,K^{-2}mol^{-1}}$.
Above $P_{\rm c}$, the field variation of $A$ is quite identical to that reported for 
CeRu$_2$Si$_2$ at $P=0.22\,{\rm GPa}$ ($H_{\rm m}\sim 11\,{\rm T}$).~\cite{Aok11_CeRu2Si2}
The weak maxima of $A$ at $H_{\rm m}$ reflects the duality between AF and FM correlations.
Above $H_{\rm m}$, good scaling of $A(H)/A(H_{\rm m})$ is observed under pressure.
As seen in Fig.~\ref{fig:Gruneisen}, 
assuming $A\sim (m^\ast)^2$ at $H_{\rm m}$,
the pressure dependence of electronic Gr\"{u}neisen parameter $\Omega (H_{\rm m}) = -\frac{\partial \ln m^\ast}{\partial \ln V}$ is quite similar between the pure system 
CeRu$_2$Si$_2$ and the Rh-doped system Ce(Ru$_{0.92}$Rh$_{0.08}$)$_2$Si$_2$.
A more accurate value of $\Omega$ determined by  thermodynamic measurements 
is known to be $\Omega (H_{\rm m}) \sim 200$ at ambient pressure in CeRu$_2$Si$_2$.
It is interesting to note that $\Omega (H_{\rm c})$ in Ce(Ru$_{0.92}$Rh$_{0.08}$)$_2$Si$_2$ strongly decreases with pressure by comparison to $\Omega (H_{\rm m})$ both in CeRu$_2$Si$_2$ and in Ce(Ru$_{0.92}$Rh$_{0.08}$)$_2$Si$_2$.

\subsection{Comparison with Ce$_{0.9}$La$_{0.1}$Ru$_2$Si$_2$}
\begin{figure}[tbh]
\begin{center}
\includegraphics[width=1 \hsize,clip]{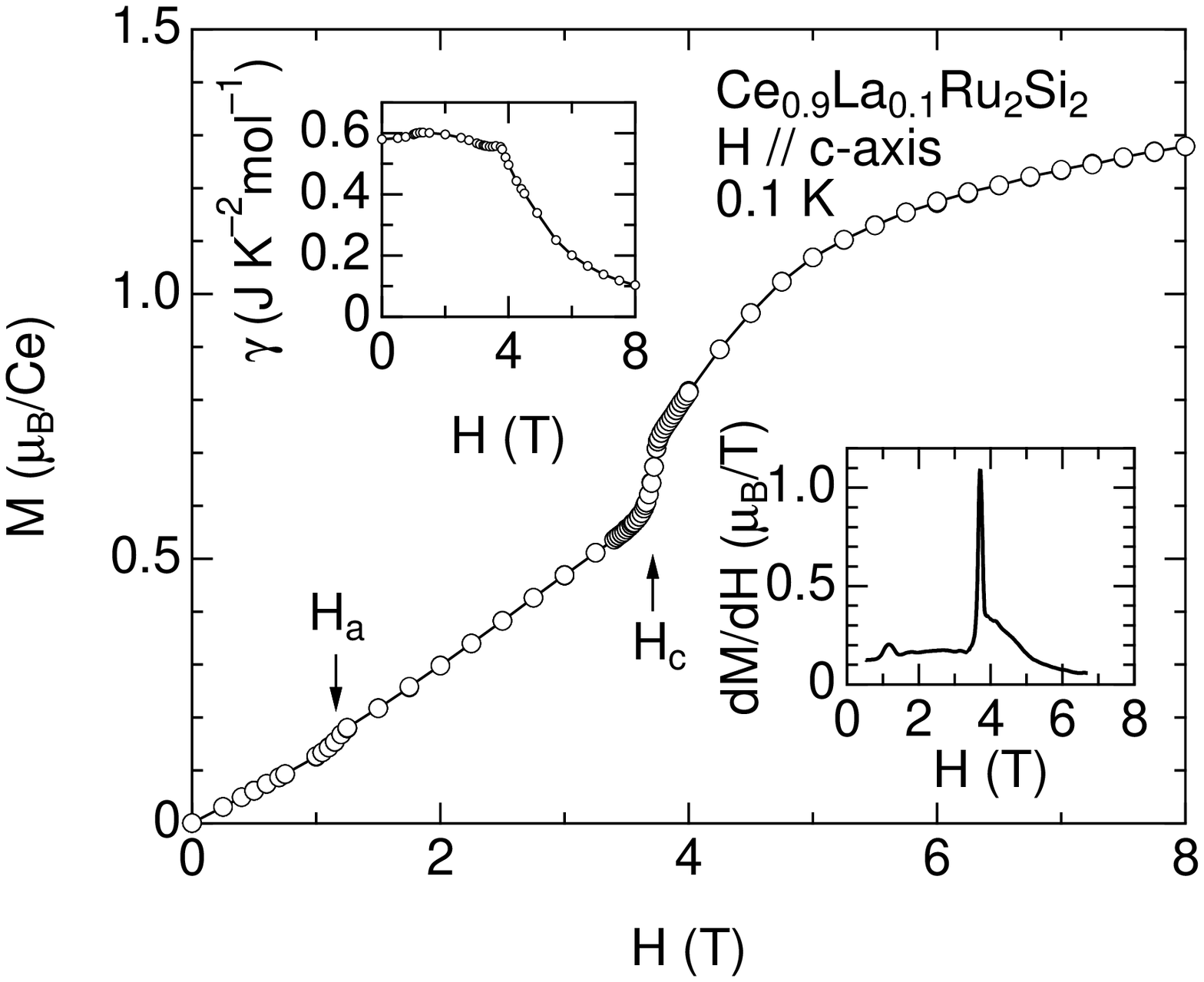}
\end{center}
\caption{Magnetization curve for $H \parallel c$-axis at $0.1\,{\rm K}$ in Ce$_{0.9}$La$_{0.1}$Ru$_2$Si$_2$. The inset at right-bottom is the field-derivative of the magnetization curve. The inset at left-top is the field dependence of the $\gamma$-value obtained from the temperature dependence of the magnetization using the Maxwell relation.}
\label{fig:La_mag}
\end{figure}

Next we compare the results of Ce(Ru$_{0.92}$Rh$_{0.08}$)$_2$Si$_2$ with those of Ce$_{0.9}$La$_{0.1}$Ru$_2$Si$_2$,
focusing on the $10\,{\%}$ doped crystal,
as it presents similar impurity scattering, as known from the value of $\rho_0$.~\cite{Hae96,Fis91}
Its N\'{e}el temperature $T_{\rm N}$ is around $2.5\,{\rm K}$ but
the specific heat anomaly at $T_{\rm N}$ is much broader than the one measured for
the previous Rh doped case, as it is closer to AF--PM instabilities.
Focus is given here on recent magnetization data published in Ref.~\citen{Aok11_CeRu2Si2}.
As seen in Fig.~\ref{fig:La_mag} for the transverse AF ordered mode,
a clear metamagnetic transition occurs at $H_{\rm c}$ without the separation between $H_{\rm c}$ and $H_{\rm m}$ at low temperatures.
However, on warming, a marked difference between $H_{\rm c}$ and $H_{\rm m}$ was already observed 
since $H_{\rm c}(T)$ decreases strongly.
The transition between the two transverse modes at $H_{\rm a}\sim 1\,{\rm T}$ is marked by a rather broad maximum of $\partial M/\partial H$.
Let us notice the large difference in shape and amplitude in $\partial M/\partial H$
detected at $H_{\rm c}$ between Ce(Ru$_{0.92}$Rh$_{0.08}$)$_2$Si$_2$ and Ce$_{0.9}$La$_{0.1}$Ru$_2$Si$_2$.
The inset of Fig.~\ref{fig:La_mag} shows the field variation of $\gamma (H)$
as derived from the Maxwell relation.
For both cases, $\gamma (H)$ shows a drastic decrease of $\gamma$ on entering in the PPM state.

\begin{figure}[tbh]
\begin{center}
\includegraphics[width=1 \hsize,clip]{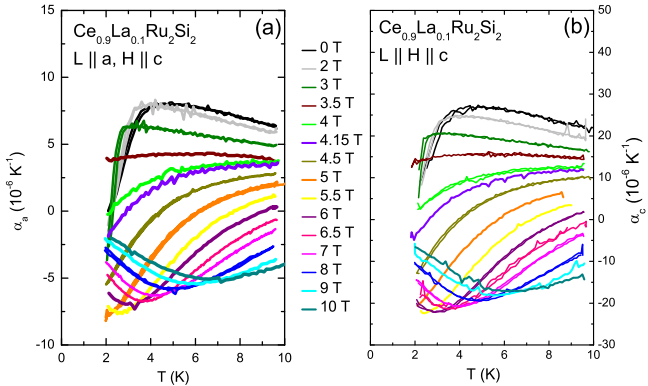}
\end{center}
\caption{(Color online) Thermal expansion of Ce$_{0.9}$La$_{0.1}$Ru$_2$Si$_2$ with high accuracy at the intermediate temperature regime ($T>2\,{\rm K}$) at different fields for $L\parallel a$-axis and $L\parallel c$-axis.}
\label{fig:La_Texp}
\end{figure}

\begin{figure}[tbh]
\begin{center}
\includegraphics[width=1 \hsize,clip]{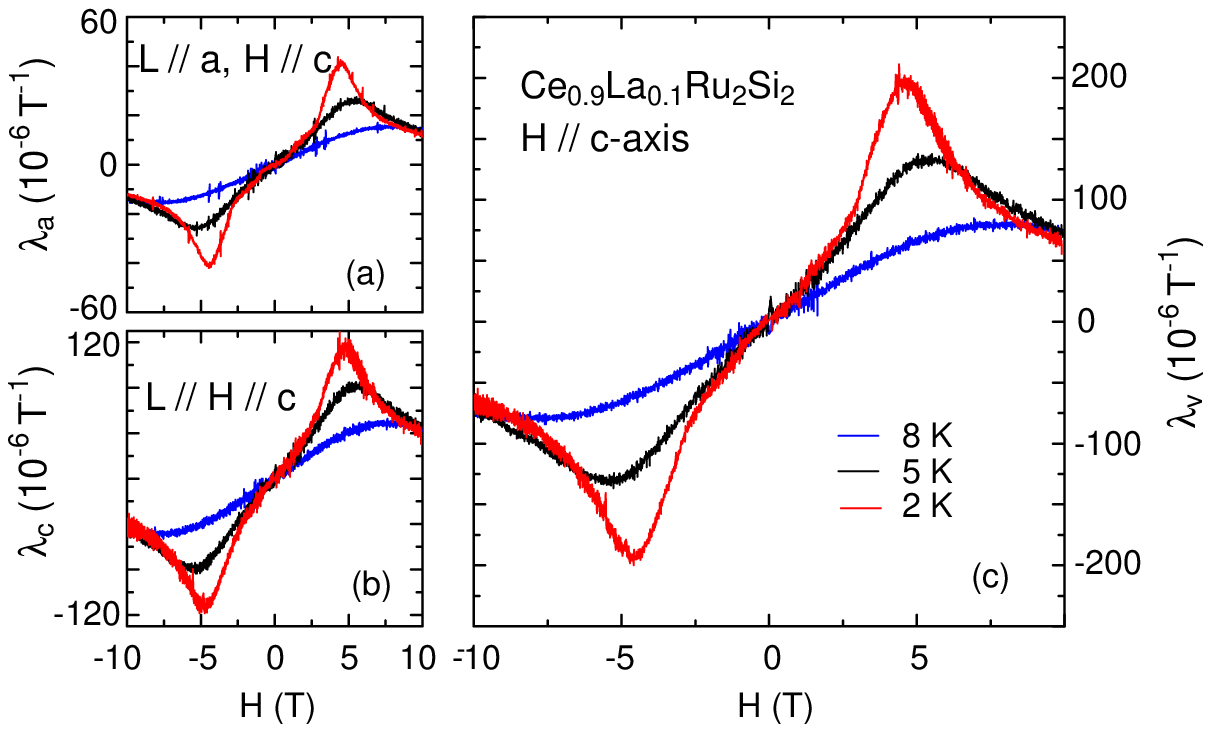}
\end{center}
\caption{(Color online) Field dependence of the magnetostriction $\lambda_V = (1/V)(\partial V/\partial H)$ on Ce$_{0.9}$La$_{0.1}$Ru$_2$Si$_2$.}
\label{fig:La_MS}
\end{figure}

\begin{figure}[tbh]
\begin{center}
\includegraphics[width=0.8 \hsize,clip]{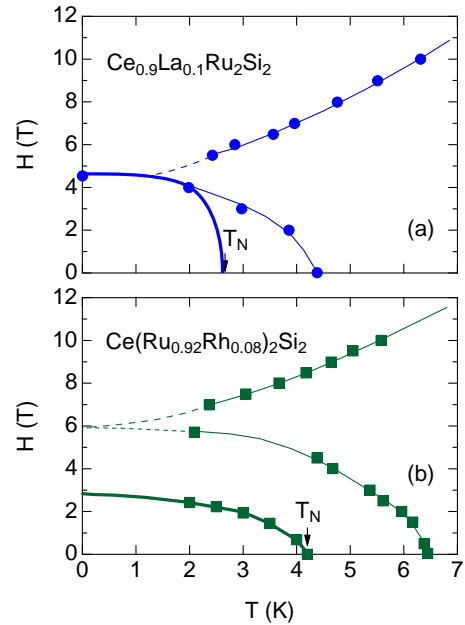}
\end{center}
\caption{(Color online) ($H,T$) phase diagram of (a) Ce$_{0.9}$La$_{0.1}$Ru$_2$Si$_2$ and (b) Ce(Ru$_{0.92}$Rh$_{0.08}$)$_2$Si$_2$ obtained by the precise thermal expansion measurements.
Lines are guides to the eyes. Thin lines correspond to the crossover from results of the temperature dependence of thermal expansion. Thick lines correspond to the AF boundary. The data of AF boundary for Ce$_{0.9}$La$_{0.1}$Ru$_2$Si$_2$ are extrapolated from the previous low temperature measurements~\cite{Fis91}.
}
\label{fig:Texp_phase_Rh_La}
\end{figure}

Precise thermal expansion measurements were also realized on Ce$_{0.9}$La$_{0.1}$Ru$_2$Si$_2$ down to $2\,{\rm K}$ (Fig.~\ref{fig:La_Texp}).
Due to the weakness of $T_{\rm N}\sim 2.5\,{\rm K}$ and also the rather broad specific heat anomaly associated to the onset of the ordering,
the AF boundary is difficult to define. However as for Ce(Ru$_{0.92}$Rh$_{0.08}$)$_2$Si$_2$, 
clear extrema of $\alpha$ in temperature allow to determine the $\tilde{T}(H)$ crossover (Fig.~\ref{fig:La_MS}).
The corresponding phase diagram is shown in Fig.~\ref{fig:Texp_phase_Rh_La}.
Quite similar behaviors emerge with the sign that AF correlations will be initially stronger for Ce(Ru$_{0.92}$Rh$_{0.08}$)$_2$Si$_2$
in good agreement with the ranking of their $T_{\rm N}$.

\section{Discussion}
\begin{figure}[tbh]
\begin{center}
\includegraphics[width=1 \hsize,clip]{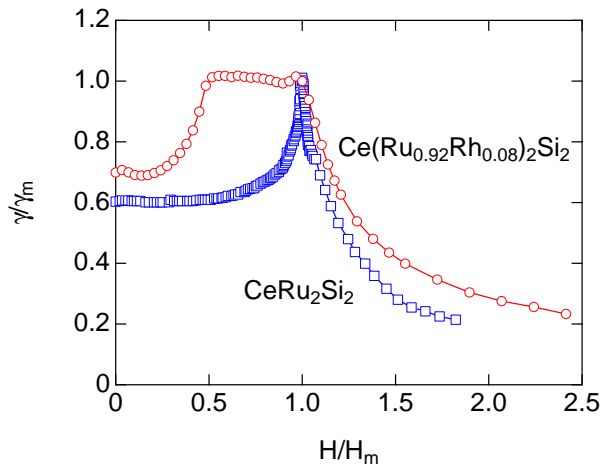}
\end{center}
\caption{(Color online) Field dependence of the Sommerfeld coefficient $\gamma$ in Ce(Ru$_{0.92}$Rh$_{0.08}$)$_2$Si$_2$ and CeRu$_2$Si$_2$. The field and $\gamma$ are scaled with $H_{\rm m}$ and the $\gamma$-value at $H_{\rm m}$, respectively. The data of CeRu$_2$Si$_2$ are cited from ref.~\protect\citen{Flo10}}
\label{fig:gamma_scale}
\end{figure}

In Ce(Ru$_{0.92}$Rh$_{0.08}$)$_2$Si$_2$, the image is that $H_{\rm c}$ marks the transition from AF to PM phases.
Increasing the magnetic field above $H_{\rm c}$ may lead to escape from AF criticality 
and thus will correspond to a strong field-decrease of $\gamma$ as proposed in Fig.~\ref{fig:phase_gamma_schematic}.
However, as observed in Ce$_{0.9}$La$_{0.1}$Ru$_2$Si$_2$ where $H_{\rm a}$ marks a switch from one AF phase to another AF phase
only by changing its wave vector,
the $\gamma$-value is basically field-invariant between $H_{\rm c}$ and $H_{\rm m}$ in Ce(Ru$_{0.92}$Rh$_{0.08}$)$_2$Si$_2$.
Furthermore its value is quite close to the maximum value of $\gamma_{\rm c}\sim 650\,{\rm mJ\,K^{-2}mol^{-1}}$
reached at $P_{\rm c}$ or at $H_{\rm c}^\ast$ in the Ce$_{1-x}$La$_x$Ru$_2$Si$_2$ family.
This suggests that above $H_{\rm c}$ in Ce(Ru$_{0.92}$Rh$_{0.08}$)$_2$Si$_2$ the low energy AF spin dynamics may be dominated by the fluctuations of the transverse mode.
However, when the polarization reaches a critical value at $H_{\rm m}$ corresponding to a critical value of magnetization $M_{\rm c}\sim 0.7\,\mu_{\rm B}$,
a strong field decrease of $\gamma$ is observed quite similar to that detected on the PPM side of CeRu$_2$Si$_2$ for $M=M_{\rm c}=0.6\,\mu_{\rm B}$.
Figure~\ref{fig:gamma_scale} shows the comparison for the relative field variation of Ce(Ru$_{0.92}$Rh$_{0.08}$)$_2$Si$_2$ and CeRu$_2$Si$_2$~\cite{Flo05,Flo02_CeRu2Si2,Flo10}
in a $\gamma/\gamma_{\rm m}$ vs $H/H_{\rm m}$ representation.
For CeRu$_2$Si$_2$ a sharp maximum occurs at $H_{\rm m}$ as the magnetic field close to $H_{\rm m}$ wipes out the AF correlations,
while the growth of uniform magnetization is associated with an increase of the FM correlations
which slows down just at the vicinity of $H_{\rm m}$.~\cite{Flo05}
Increasing further the magnetic field above $H_{\rm m}$ reinforces the local character and thus
leads to recover the properties of polarized Kondo centers.

Previous results on Ce(Ru$_{0.85}$Rh$_{0.15}$)$_2$Si$_2$
($T_{\rm N}=5.5\,{\rm K}$, $\Vec{k}_3=0.42$, $H_{\rm c}=3.5\,{\rm T}$, $H_{\rm m}=5\,{\rm T}$)
give an extrapolation of $\gamma (H)$ with only a rounded maxima near $H_{\rm m}$ ($\gamma(H_{\rm m})=550\,{\rm mJ\,K^{-2}mol^{-1}}$,
while the zero field value reaches $\gamma(0)\sim 300\,{\rm mJ\,K^{-2}mol^{-1}}$;
those two values are quite close to the ones found in the present experiment for $x=0.08$.
Approaching $x_{\rm c}$ leads to an increase of the $H_{\rm c}$--$H_{\rm m}$ window (by a factor 2)
and thus to an excellent decoupling between AF and pseudo-metamagnetic instabilities.
As mentioned later, low doping allows also to minimize the effect of lattice mismatch.

The novelty in the Rh doped case is the mismatch in the lattice parameters of CeRu$_2$Si$_2$ and CeRh$_2$Si$_2$, which is tabulated in Table~\ref{tab:}
together with the comparison with LaRu$_2$Si$_2$.
In La-doped case, both $a$ and $c$ lattice parameters expand.
The driving force is the volume and the local perturbation is only moderated.
In Rh-doped case, the value of $a$ decreases, while the value of $c$ increases.
This contradictory behavior boosts the local perturbation now not reduced to sole volume effects.
Furthermore Fermi level may change by the substitution of Ru by Rh as well as the local fluctuations
due to an increase of the carrier number of $d$ electrons.
The volume change between CeRu$_2$Si$_2$ and CeRh$_2$Si$_2$ is only $4\times 10^{-3}$ 
corresponding to a compression by $4\,{\rm kbar}$.
Despite this volume reduction, $T_{\rm N}$ in CeRh$_2$Si$_2$ reaches $36\,{\rm K}$ and $H_{\rm c}$ is $26\,{\rm T}$.
Furthermore AF occurs for a commensurate wave vector $(1/2, 1/2, 0)$ quite different from $(0,0,0.35)$
of Rh-doped CeRu$_2$Si$_2$.
Estimation of the Kondo temperature $T_{\rm K}\sim 50\,{\rm K}$ for CeRh$_2$Si$_2$
instead of $T_{\rm K}\sim 25\,{\rm K}$ for CeRu$_2$Si$_2$.
In the popular image of the Doniach collapse of AF for a Kondo lattice, the Ruderman--Kittel--Kasuya--Yosida (RKKY) interactions in CeRh$_2$Si$_2$ must be quite larger than that of CeRu$_2$Si$_2$.
Thus a source of competing interactions is created via the Rh substitution,
plus additional change in the carrier number caused by the Rh doping.
At least an important consequence is that at $H=0$ the magnetic ordered mode became the longitudinal one.
It was recently observed that slight changes of the Fermi surface occur on doping with La or Ge substitution.~\cite{Mat10_CeRu2Si2,Mat11_CeRu2Si2}
It seems indirectly here that Rh doping has more drastic effect which promotes the $\Vec{k}_3$ instabilities.

Evidences of strong nesting properties with partial gap opening for the Rh-doped system were given in the resistivity measurements,
when the current $J$ is applied along the $c$-axis as shown in Fig.~\ref{fig:resist}(a)(b).
For Ce(Ru$_{0.85}$Rh$_{0.15}$)$_2$Si$_2$, similar data can be found in ref.~\citen{Mur97}.
No jump of resistivity for $J \parallel c$-axis was observed in the La-doped case,
while with Ge-doping, the nesting is detected for $J\parallel a$-axis~\cite{Ama11}.
We have verified that the nesting persists also in the present system Ce(Ru$_{0.92}$Rh$_{0.08}$)$_2$Si$_2$,
as shown in Fig.~\ref{fig:resist}(b).
\begin{table}[tbhp]
\caption{Lattice parameters and volumes of CeRu$_2$Si$_2$, LaRu$_2$Si$_2$ and CeRh$_2$Si$_2$.}
\label{tab:}
\begin{tabular}{lccc}
\hline
			& CeRu$_2$Si$_2$		& LaRu$_2$Si$_2$		& CeRh$_2$Si$_2$		\\
\hline
$a$			& $4.192\,{\rm \AA}$	& $4.215\,{\rm \AA}$	& $4.09\,{\rm \AA}$	\\
$c$			& $9.78\,{\rm \AA}$	& $9.93\,{\rm \AA}$	& $10.18\,{\rm \AA}$	\\
$c/a$		& 2.32				& 2.20				& 2.48				\\
$V$			& $171\,{\rm \AA}^3$	& $176.4\,{\rm \AA}^3$	& $170.3\,{\rm \AA}^3$	\\
\hline
\end{tabular}
\end{table}

A difficult enigma is what would be the phase diagram of a pure lattice of CeRu$_2$Si$_2$ (without induced disorder),
if the ground state at $H=0$ would be an AF longitudinal mode;
would the transition from AF to PM at $H_{\rm c}$ be replaced by a switch between two AF structures 
as observed for the La substitution?
As pointed out in Fig.~\ref{fig:phase_schematic},
the AF regime~\cite{Mig91,Mig90} is rather complex at $T_{\rm N}(H=0)$.
A new question is if, due to the Rh substitution, above $H_{\rm c}$, the system will switch from AF phase dominated by a longitudinal mode to a PM nearly AF phase dominated by a transverse mode between $H_{\rm c}$ and $H_{\rm m}$ before becoming governed by the crossover to the PPM phase.
A strong indication is given by the pressure resistivity measurements suggesting quite different critical  values of $\gamma_{\rm c}$ for the longitudinal and transversal instability.

Obviously, an open question is the interplay between the different modes on the spin dynamics.
It will require a new generation of inelastic experiments,
in which it is expected that the transverse mode may govern the spin dynamics above $H_{\rm c}$.
The main difficulty is to grow a large homogenous Rh-doped single crystal,
as the physical properties are very sensitive to the development of concentration gradient.
Our choice will be to remain in the AF domain with $x\sim 0.08$
as we can verify the homogeneity from the size and shape of the specific heat anomaly at $T_{\rm N}$ for $H=0$.
Another microscopic probe will be to succeed in detecting the Fermi surfaces.  
Here application of the magnetic field along the basal plane 
gives some hope to track the Fermi surface change,
despite the fact that the electronic mean free path could be damped by doping.

Some of the figures with $H_{\rm m}$--$H_{\rm c}$ window are reminiscent of the effects observed in the cases such as Sr$_3$Ru$_2$O$_7$~\cite{Fra10} or recently UCoAl~\cite{Aok11_UCoAl}.
In Sr$_3$Ru$_2$O$_7$ it was proposed that a nematic phase appears due to the strength of spin-orbit coupling, and that weak disorder preempts this establishment at the profit of a smearing into  a metamagnetic quantum critical point.
In UCoAl, a plateau of $\gamma (H)$ was recently detected above the quantum critical endpoint.
At least, strong evidences are given here that the large $H$ plateau of constant $\gamma$ is the result of the competition between AF instability and field-driven FM instability.

\section{Conclusion}
Doping CeRu$_2$Si$_2$ with Rh instead of La or Ge leads to a drastic change of $(H,T)$ phase diagram 
of the AF phases.
The achievement of very low temperature allowed to characterize the field variation of $\gamma$ 
inside the AF domain below $H_{\rm c}=2.8\,{\rm T}$ 
inside the ``unstable'' PM phase between $H_{\rm c}$ and $H_{\rm m}=5.8\,{\rm T}$ and in the PPM phase.
The pressure study suggested that two critical values of $\gamma$ are associated respectively to longitudinal ($\Vec{k}_3$) and transversal AF instability.
A key ingredient is the lattice mismatch between CeRu$_2$Si$_2$ and CeRh$_2$Si$_2$.
Above $H_{\rm c}$, the occurrence of a PM ground state with Rh-doping is in excellent agreement with the lattice contraction by comparison to the pure system CeRu$_2$Si$_2$.

\section*{Acknowledgements}
We thank Y. Matsumoto and Y. Machida for useful discussions.
This work was supported by ERC starting grant (NewHeavyFermion), French ANR project (CORMAT, SINUS, DELICE).



\begin{thebibliography}{10}
\bibitem{Flo05}
J.~Flouquet: {\em Progress in Low Temperature Physics} (Amsterdam, 2005)
  Vol.~15, p.~139.

\bibitem{Loh07}
H.~v.~L{\"o}hneysen, A.~Rosch, M.~Vojta and P.~W{\"o}lfle: Rev. Mod. Phys. {\bf
  79} (2007) 1015.

\bibitem{Hol95}
S.~Holtmeier, P.~Haen, A.~Lacerda, P.~Lejay, J.~L. Tholence, J.~Voiron and
  J.~Flouquet: Physica B (1995) 250.

\bibitem{Flo02_CeRu2Si2}
J.~Flouquet, P.~Haen, S.~Raymond, D.~Aoki and G.~Knebel: Physica B {\bf
  319} (2002) 251.

\bibitem{Hae96}
P.~Haen, F.~Lapierre, J.~Voiron and J.~Flouquet: J. Phys. Soc. Jpn. Suppl. B
  {\bf 65} (1996) 27.

\bibitem{Que88}
S.~Quezel, P.~Burlet, J.~L. Jacoud, L.~P. Regnault, J.~Rossat-Mignod,
  C.~Vettier, P.~Lejay and J.~Flouquet: J. Magn. Magn. Mater. {\bf 76-77}
  (1988) 403.

\bibitem{Flo10}
J.~Flouquet, D.~Aoki, W.~Knafo, G.~Knebel, T.~D. Matsuda, S.~Raymond,
  C.~Proust, C.~Paulsen and P.~Haen: J. Low Temp. Phys. {\bf 161} (2010) 83.

\bibitem{Wei10}
F.~Weickert, M.~Brando, F.~Steglich, P.~Gegenwart and M.~Garst: Phys. Rev. B
  {\bf 81} (2010) 134438.

\bibitem{Hae99}
P.~Haen, H.~Bioud and T.~Fukuhara: Physica B {\bf 259-261} (1999) 85.

\bibitem{AokiH95}
H.~Aoki, M.~Takashita, S.~Uji, T.~Terashima, K.~Maezawa, R.~Settai and
  Y.~\={O}nuki: Physica B {\bf 206{\&}207} (1995) 26.

\bibitem{Jul94}
S.~Julian, F.~Tautz, G.~McMullan and G.~Lonzarich: Physica B {\bf 199-200}
  (1994) 63.

\bibitem{Dao06}
R.~Daou, C.~Bergemann and S.~R. Julian: Phys. Rev. Lett. {\bf 96} (2006)
  026401.

\bibitem{Miy06}
K.~Miyake and H.~Ikeda: J. Phys. Soc. Jpn. {\bf 75} (2006) 033704.

\bibitem{Lac89}
A.~Lacerda, A.~de~Visser, L.~Puech, P.~Lejay, P.~Haen, J.~Flouquet, J.~Voiron
  and F.~J. Okhawa: Phys. Rev. B {\bf 40} (1989) 11429.

\bibitem{Pau90}
C.~Paulsen, A.~Lacerda, L.~Puech, P.~Haen, P.~Lejay, J.~L. Tholence,
  J.~Flouquet and A.~{de Visser}: J. Low Temp. Phys. {\bf 81} (1990) 317.

\bibitem{Ros88}
Rossat-Mignot, L.~P. Regnault, J.~L. Jacoud, C.~Vettier, P.~Lejay, J.~Flouquet,
  E.~Walker, D.~Jaccard and A.~Amato: J. Magn. Magn. Mater. {\bf 76-77} (1988)
  376.

\bibitem{Flo04}
J.~Flouquet, Y.~Haga, P.~Haen, D.~Braithwaite, G.~Knebel, S.~Raymond and
  S.~Kambe: J. Magn. Magn. Mater. {\bf 272-276} (2004) 27.

\bibitem{Sat04}
M.~Sato, Y.~Koike, S.~Katano, N.~Metoki, H.~Kadowaki and S.~Kawarazaki: J.
  Phys. Soc. Jpn. {\bf 73} (2004) 3418.

\bibitem{Kad04}
H.~Kadowaki, M.~Sato and S.~Kawarazaki: Phys. Rev. Lett. {\bf 92} (2004)
  097204.

\bibitem{Mig91}
J.~M. Mignot, L.~P. Regnault, J.~L. Jacoud, J.~Rossat-Mignod, P.~Haen and
  P.~Lejay: Physica B {\bf 171} (1991) 357.

\bibitem{Mig90}
J.-M. Mignot, J.-L. Jacoud, L.-P. Regnault, J.~Rossat-Mignod, P.~Haen,
  P.~Lejay, P.~Boutrouille, B.~Hennion and D.~Petitgrand: Physica B {\bf
  163} (1990) 611.

\bibitem{Sek92}
C.~Sekine, T.~Sakakibara, H.~Amitsuka and Y.~M. Goto: J. Phys. Soc. Jpn. {\bf
  61} (1992) 4536.

\bibitem{Sek93}
C.~Sekine, T.~Yoshida, S.~Murayama, K.~Hoshi and T.~Sakakibara: Physica B {\bf
  186-188} (1993) 511.

\bibitem{Sak92}
T.~Sakakibara, C.~Sekine, H.~Amitsuka and Y.~Miyako: J. Magn. Magn. Mater. {\bf
  108} (1992) 193.

\bibitem{Kaw97}
S.~Kawarazaki, M.~Sato, H.~Kadowaki, Y.~Yamamoto and Y.~Miyako: J. Phys. Soc.
  Jpn. {\bf 66} (1997) 2473.

\bibitem{Kad06}
H.~Kadowaki, Y.~Tabata, M.~Sato, N.~Aso, S.~Raymond and S.~Kawarazaki: Phys.
  Rev. Lett. {\bf 96} (2006) 016401.

\bibitem{Sek97_CeRu2Si2}
C.~Sekine, Y.~Nakazawa, K.~Kanoda, T.~Sakakibara, S.~Murayama, I.~Shirotani and
  Y.~\={O}nuki: Physica B {\bf 230-232} (1997) 172.

\bibitem{Sek98}
C.~Sekine, T.~Tayama, T.~Sakakibara, S.~Murayama, I.~Shirotani and
  Y.~\={O}nuki: J. Magn. Magn. Mater. {\bf 177-181} (1998) 411.

\bibitem{Pau11}
C.~Paulsen, D.~Aoki, G.~Knebel and J.~Flouquet: J. Phys. Soc. Jpn. {\bf 80}
  (2011) 053701.

\bibitem{Fis91}
R.~A. Fisher, C.~Marcenat, N.~E. Phillips, P.~Haen, F.~Lapierre, P.~Lejay,
  J.~Flouquet and J.~Voiron: J. Low Temp. Phys. {\bf 84} (1991) 49.

\bibitem{Aok11_CeRu2Si2}
D.~Aoki, C.~Paulsen, T.~D. Matsuda, L.~Malone, G.~Knebel, P.~Haen, P.~Lejay,
  R.~Settai, Y.~\={O}nuki and J.~Flouquet: J. Phys. Soc. Jpn. {\bf 80} (2011)
  053702.

\bibitem{Bio99}
H.~Bioud: PhD thesis (Grenoble, 1998)

\bibitem{Hae01}
P.~Haen, S.~Kambe, H.~Bioud and A.~{de Visser}: J. Magn. Magn. Mater. {\bf
  226-230} (2001) 252.

\bibitem{Lac88}
A.~Lacerda: PhD thesis (Grenoble, 1990).

\bibitem{Kam97}
S.~Kambe and J.~Flouquet: Solid State Commun. {\bf 103} (1997) 551.

\bibitem{Mat11_CeRu2Si2}
Y.~Matsumoto, M.~Sugi, K.~Aoki, Y.~Shimizu, N.~Kimura, T.~Komatsubara, H.~Aoki,
  M.~Kimata, T.~Terashima and S.~Uji: J. Phys. Soc. Jpn. {\bf 80} (2011)
  074715.

\bibitem{Pue88}
L.~Puech, J.~M. Mignot, P.~Lejay, P.~Haen, J.~Flouquet and J.~Voiron: J. Low
  Temp. Phys. {\bf 70} (1988) 237.

\bibitem{Mat10_CeRu2Si2}
Y.~Matsumoto, N.~Kimura, H.~Aoki, M.~Kimata, T.~Terashima, S.~Uji, T.~Okane and
  H.~Yamagami: J. Phys. Soc. Jpn. {\bf 79} (2010) 083706.

\bibitem{Mur97}
S.~Murayama, C.~Sekine, A.~Yokoyanagi, K.~Hoshi and Y.~{\=O}nuki: Phys. Rev. B
  {\bf 56} (1997) 11092.

\bibitem{Ama11} 
Y.~Amakai, E.~Harada, D.~Yokoyama, S.~Murayama, K.~Matsumoto, H.~Takano, N.~Momono, K.~Matsubayashi and Y.~Uwatoko: 
J. Phys. Soc. Jpn. {\bf 80} (2011) SA062.

\bibitem{Fra10}
E.~Fradkin, S.~A. Kivelson, M.~J. Lawler, J.~P. Eisenstein and A.~P. Mackenzie:
  Annual Rev. Cond. Mat. Phys. {\bf 1} (2010) 153.

\bibitem{Aok11_UCoAl}
D.~Aoki, T.~Combier, V.~Taufour, T.~D. Matsuda, G.~Knebel, H.~Kotegawa and
  J.~Flouquet: J. Phys. Soc. Jpn. {\bf 80} (2011) 094711.

\end{thebibliography}

\end{document}